# A quantitative fusion strategy of stock picking and timing based on Particle Swarm Optimized-Back Propagation Neural Network and Multivariate Gaussian-Hidden Markov Model

HUAJIAN, LI*, Guangzhou University, Guangzhou, Guangdong, China, 2112005065@e.gzhu.edu.cn

LONGJIAN, LI, Guangdong University of Technology, Guangzhou, Guangdong, China, 2112109116@mail2.gdut.edu.cn

JIAJIAN, LIANG, Guangzhou University, Guangzhou, Guangdong, China, 1112315014@e.gzhu.edu.cn

WEINAN, DAI, Trine University, Phoenix, Arizona, USA, wdai22@my.trine.edu

In recent years, machine learning (ML) has brought effective approaches and novel techniques to economic decision, investment forecasting, and risk management, etc., coping the variable and intricate nature of economic and financial environments. For the investment in stock market, this research introduces a pioneering quantitative fusion model combining stock timing and picking strategy by leveraging the Multivariate Gaussian-Hidden Markov Model (MGHMM) and Back Propagation Neural Network optimized by Particle Swarm (PSO-BPNN). After the information coefficients (IC) between fifty-two factors that have been winsorized, neutralized and standardized and the return of CSI 300 index are calculated, a given amount of factors that rank ahead are choose to be candidate factors heading for the input of PSO-BPNN after dimension reduction by Principal Component Analysis (PCA), followed by a certain amount of constituent stocks outputted. Subsequently, we conduct the prediction and trading on the basis of the screening stocks and stock market state outputted by MGHMM trained using inputting CSI 300 index data after Box-Cox transformation, bespeaking eximious performance during the period of past four years. Ultimately, some conventional forecast and trading methods are compared with our strategy in Chinese stock market. Our fusion strategy incorporating stock picking and timing presented in this article provide a innovative technique for financial analysis.

## 1 INTRODUCTION

More and more attention has been dedicated to the analysis and computation of Quantitative finance since Modern Portfolio Theory was proposed in 1952 by Harry Markowitz [1]. Amongst the previous research, numerous scholars made endeavour to the forecast and investment with regard to miscellaneous financial objects such as bond, stock, option, future, etc. In particular, plenty of quantitative methods [2-5] emerged as a tool to make the trade decisions about avoiding risk and delving in stock market for the excess return of investment portfolios through mathematical models and computer technology. Opposite from traditional qualitative methods which rely on experience, sensation or emotion to make judgments, quantitative methods are impervious to subjective consciousness. In recent years, the advancement of artificial intelligence and machine learning has provided a expansive and effective path for the study of quantitative finance.

---

* Corresponding author.

However, myriad factors that could affect the stock market caused its fluctuation and intricateness, like national policy, business cycles, behavior of investor and enterprise, and so forth. Hence almost all conventional strategies have their limitations. Screening the promising stocks (picking) and selecting the appropriate time (timing) to trade are two major challenges. Hitherto, researches integrating picking and timing are greatly scarce. In this paper, we propose a fusion strategy integrating picking and timing, which is a remarkable method for investors in stock market.

Neural Network (NN) is a kind of nonlinear information processing system imitating the architecture and function of cerebral nerves whose rationale takes root in biological neurology and signal transmission mechanism. In short, Neural Network is a mathematical model of artificial brain. It is not necessary to determine the mapping relationship between input and output in advance for Neural Network, merely through training and learning, capable of deriving the results adjacent to the authentic mapping when given the input. Nowadays, Neural Network has been applied in wide realm such as chemistry [6], industry [7], medical detection [8], economic forecasting [9], etc.

Hidden Markov Model (HMM) is a probabilistic model of time series, born for depicting a Markov process with latent variables. After learning and training, the parameters and latent variables of HMM are able to be estimated and determined in accordance with a sequence of observable variables. It is successfully employed in speech recognition [10], natural language processing [11], bioinformatics [12], etc.

In this article we present a quantitative fusion strategy integrating picking and timing which blend PSO-BPNN and MGHMM. Amid our ensemble approach, the whole trial was conducted in CSI 300 index market. Firstly, we figure out the information coefficients between fifty-two factors whose sequence data have been winsorized, neutralized and standardized and the return of CSI 300 index. Subsequently a certain amount of factors which come out in front are screened out to be candidates as the input of Back Propagation Neural Network (BPNN) following dimension reduction by PCA. In which we utilize the Particle Swarm algorithm with its merits of swifter and more effortless global convergence to optimize the BPNN. BPNN after training is able to estimate the returns of all constituent stocks and output $n$ maximum stocks with respect to predictive return for subsequent trade. Five factors of CSI 300 index are selected to be the observable variables in HMM which are multidimensional vector assumed as following Multivariate Gaussian distribution. Every observable variable sequence thereby undergo Box-Cox transformation in advance. MGHMM after training is competent to output the hidden states that indicate distinct market states such as rise, fall, oscillation, etc. At length, we carry out the backtesting according to the outputted hidden states inside a stock pool consist of $n$ stocks outputted by PSO-BPNN in the historical data from the past four years, and compare our fusion strategy with several conventional methods in A-Share market.

## 2 RELATED WORK

The advent of big data era eventuated in speedy development of quantitative finance, wherein including a substantial proportion of machine learning techniques as mighty tools for financial analysis. In what follows a concise review of some key researches about the application of NN and HMM in financial market are presented.

## 2.1 Studies about NN

NN renowned for its formidable non-linear analysis ability, heretofore is extensively applied in financial market. White [13] has pioneered the employment of NN to predict the daily return of IBM stock, giving rise to the prevalence of NN in financial market. Based on stock prediction principle of BPNN, Li et al. [14] drew on three-layered feed forward neural network to construct a forecast model for stock market taking advantage of Levenberg-Marquardt BP algorithm, and the results of simulation experiment for the typical index from Shanghai stock exchange market illustrated the model could achieve effective short-term forecast. Inthachot et al. [15] used NN and Support Vector Machine (SVM) simultaneously to forecast the trend of SET50 index in Thailand's emerging stock market during a period of 2009 to 2013, indicating that the prediction precision of NN is significantly higher than that of SVM. Qiu et al. [16] applied NN optimized by Genetic algorithms (GA) in predicting the movement direction of morrow price of the Japanese stock market index, the experimental results demonstrated the more exceptional predictability and performance as to the prediction task in contrast with prior researches. Fitriyaningsih et al. [17] predict the closing price of S&P 500 stock exchange putting historical data composed of five variables as input of BPNN. The results exhibited a superior performance with the Mean Average Percentage Error (MAPE) of 0.2307 and a high prediction accuracy of 99.77% for a month. Goel et al. [18] took some macroeconomic indicators as input variables, used NN optimized by Scaled Conjugate Gradient Algorithm (SCG) to portray the Indian stock market, and the empirical results manifested that SCG-NN model could attain a accuracy of 93% in forecasting the closing prices of Bombay Stock Exchange (BSE) Sensex. Wang et al. [19] harnessed NN to forecast the stock price crashes of Chinese equity market, and found that the NN with fewer hidden layers surpass the NN with more ones, besides outperforming the classical logistic regression model as well. Zeng [20] synergized Convolutional Neural Networks (CNN) and Transformer to predict intraday price variation of S&P 500 constituents, the empirical outcomes bespoke that the proposed model outstripped some usual statistical approaches.

## 2.2 Studies about HMM

HMM is endowed with potent analysis capacity for time series data, which is explainable with firm statistical foundation, ergo it is suitable quite for examining financial market behavior. Hassan and Nath [21] first applied HMM to financial market. The trained HMM was utilized to search the stock price behavior pattern that was the identical or approximate to today's from historical data to predict the airlines stock price. By comparison, the prediction accuracy of proposed model was not inferior to that of NN. Thereafter, Hassan and Nath [22] proposed an improved fusion algorithm. The daily observable variables were converted into the data series with higher resolution by NN and then serve as the input of HMM. GA was employed to optimize the initial parameters of HMM instead of Baum-Welch algorithm (BWA) so as to promote the ability of global optimization. The empirical conclusion displayed that the proposed amelioration is more accurate than before. Nguyen [23] used Bayesian information (BIC), Consistent Akaike Information (CAIC), Akaike information (AIC) and Hannan Quinn information (HQC) to deduce the optimum hidden states number of HMM. After training, HMM is wielded to forecast and trade the monthly price of S&P 500 index emanating the superior deed than HAR model. Zhang et al. [24] presented a dynamic high-order HMM approach to investigate the relevance between outputted hidden states and the price alteration trend of CSI 300 and S&P 500 stock index by reducing the high-dimensional state vector into one dimension. Experiments proved that the provided high-order HMM has preferable competence to forecast market trend than the first-order one and lower risk in trading. Lolea et al. [25] applied HMM to construct trading strategy during the COVID-19 pandemic period, and the trading trial showed that retail investors could gain a higher excess return than the market based on the signal provided by HMM during the turbulent period. Given the homogeneous Markov hypothesis of first-order HMM is unreasonable, that is, the state at current moment is only related to that at last moment. Therefore, the first-order continuous HMM was extended to the second-order one by Zhi Su and Bo Yi [26] and used to predict stock price combining the fluctuation range measure. The simulation on Hang Seng Index (HSI) showed that the proposed model exhibited a remarkable prediction precision.

## 3 METHODOLOGY

### 3.1 PSO-BPNN for stock picking

#### 3.1.1 Overview of BPNN

BPNN is a multi-layer feedforward neural network trained with the error backward propagation algorithm. There are input layer, hidden layers and output layer in BPNN. The BPNN used in this paper has one input layer with a undetermined number of nodes, one hidden layer with a undetermined number of nodes and one output layer with one nodes, whose architecture can be seen in Figure 1.

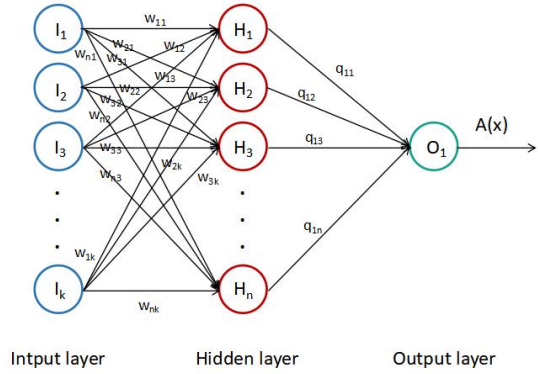

Figure 1: BPNN architecture.

In Figure 1, $I_i$, $i = 1, 2, \cdots, k$, represent the value of each node in input layer; $H_i$, $i = 1, 2, \cdots, n$, represent the bias value of each node in hidden layer; $O_1$ is the bias value in output layer; $w_{ij}$, $i = 1, 2, \cdots, n$; $j = 1, 2, \cdots, k$, are the weight from input layer to hidden layer; and $q_{1i}$, $i = 1, 2, \cdots, n$, is the weight from hidden layer to output layer; $A(x)$ is the activation function which is depicted as equation (1).

$$A(x) = 0.1 \times \frac{e^{ax} - 1}{e^{ax} + 1} \tag{1}$$

Where $a$ is a pending coefficient. First, the training sample enters the BPNN through the input layer, then passes through the neurons in hidden layer after weight assignment, next the whole summation accesses output layer as the output. The aforementioned process is the forward propagation of BPNN. The output value can be calculated by equation (2).

$$Output = A[\sum_{i=1}^{n} q_{1j}(\sum_{j=1}^{k} w_{ij}I_j + H_i) + O_1] \tag{2}$$

In the training process, when the output value differs significantly from the actual one with a large error outweighing a threshold, the error signal is returned backward along the original connection path, the weight and bias value in BPNN will be modified. This process is the back propagation of BPNN. In this way, the parameters inside BPNN shall be upgraded and optimized repeatedly, so that the output error is gradually reduced until below the threshold and the desired value is approached.

*3.1.2 Overview of PSO*

We employ PSO, which originates from studies on the predation behavior of bird flocks, to optimize BPNN owing to its speedy and puissant global convergence ability. The iteration flowchart of PSO can be seen in Figure 2.

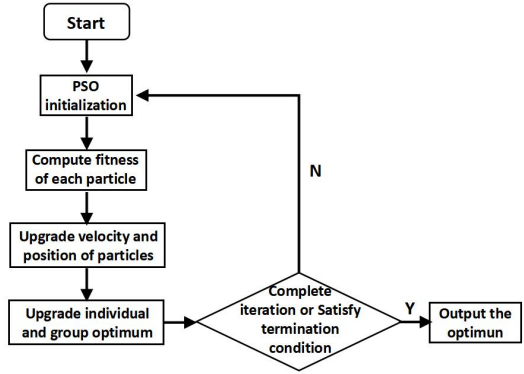

Figure 2: PSO iteration flowchart.

For the BPNN with $M$ parameters, the formula for updating velocity $\vec{X}_i^n = (x_{i1}, x_{i2}, \cdots, x_{iM})^T$ and position $\vec{V}_i^n = (V_{i1}, V_{i2}, \cdots, V_{iM})^T$ of the *i-th* particle can be calculated by equation (3) and (4).

$$\begin{cases} \vec{V}_i^{n+1} = C\vec{V}_i^n + r_1 c_1 (\vec{I}_i^n - \vec{X}_i^n) + r_2 c_2 (\vec{G}_i^n - \vec{X}_i^n) \\ \vec{X}_i^{n+1} = \vec{X}_i^n + \vec{V}_i^{n+1} \end{cases} \tag{3}$$

Where the superscript *n* and *n+1* refer to the *n-th* and *n+1-th* iteration; $c_1$ and $c_2$ are two acceleration coefficient; $r_1, r_2 \in [0,1]$ are generated randomly; $\vec{I}_i^n$ and $\vec{G}_i^n$ are the individual and group optimal position of the *i-th* particle; $C$ is the time-variant inertia weight described as equation (4).

$$C = W_{max} - \frac{W_{max} - W_{min}}{n} \tag{4}$$

Where $W_{max}$ and $W_{min}$ are the initial inertia weight and the ultimate inertia weight respectively. The fitness of the *i-th* particle is computed with Root Mean Square Error (RMSE) as equation (5).

$$Fitness_i = \sqrt{\frac{1}{d}\sum_{i=1}^{d}(Output_i - R_i^{next})^2} \qquad (5)$$

Where $d$ is the number of daily bar data; $R_i^{next}$ represents the actual return the next trade day. Furthermore, the PSO parameters setup is shown in Table 1.

Table 1: The parameters setup of PSO

| Parameters | Value | Meaning |
|---|---|---|
| $c_1$ | 1.5 | Acceleration coefficient one |
| $c_2$ | 1.5 | Acceleration coefficient two |
| $W_{max}$ | 0.9 | Initial inertia weight |
| $W_{min}$ | 0.4 | Ultimate inertia weight |
| $I_{max}$ | 300 | Maximum iterations |
| PS | 100 | Population size |
| $P_{max}$ | 3 | Maximum position boundary |
| $P_{min}$ | -3 | Minimum position boundary |
| $V_{max}$ | 0.1 | Maximum velocity boundary |
| $V_{min}$ | -0.1 | Minimum velocity boundary |
| AMP | 0.2 | Adaptive mutation probability |

### 3.1.3 Data preprocessing

Fifty-two candidate factors of CSI 300 index constituent stocks shown in Appendix is selected for stock picking with BPNN. Raw data are sourced from Uqer database that will undergo an meticulous preprocessing regime as follows before inputting BPNN:

1. Delete the entire rows where there are null values locating.
2. Eliminate the extrema by $3\sigma$ method.
3. Neutralize factors data in regard to market capitalization and industry.
4. Standardize factors data by *Z-score* method.
5. Calculate the IC (Pearson Correlation Coefficient) between all factors of CSI 300 index constituent stocks and the next trade day's return of CSI 300 index and sort them according to their absolute size, then select a pending number of the preceding factors.
6. Reduce the dimension of factors data to a pending dimension by PCA algorithm.

Above procedure is shown in Figure 5.

### 3.1.4 Structural optimization of BPNN

We chose the data from 2022-6-1 to 2022-6-8 as training set and the one from 2022-6-9 to 2022-6-16 as testing set to optimize the hyper-parameters comprising the hidden nodes number, the input nodes number, the coefficient $a$ of $A(x)$ and the number of screen factors under the setup presented in Table 1 through control variate method. The RMSE and Mean Absolute Error (MAE) calculated by equation (6) is chosen to evaluate the optimization process and the results is revealed in Figure 3.

$$\begin{cases} RMSE = \sqrt{\dfrac{1}{nd}\sum_{t=1}^{d}\sum_{i=1}^{n}(Prediction_i^t - R_i^{t+1})^2} \\ MAE = \dfrac{1}{nd}\sum_{t=1}^{d}\sum_{i=1}^{n}|Prediction_i^t - R_i^{t+1}| \end{cases} \quad (6)$$

Where $Prediction_i^t$ is the *i-th* stock's return at date *t*; $R_i^{t+1}$ is the CSI 300 index's return at date *t+1*; *n* and *d* refer to the stocks number and the dates number respectively.

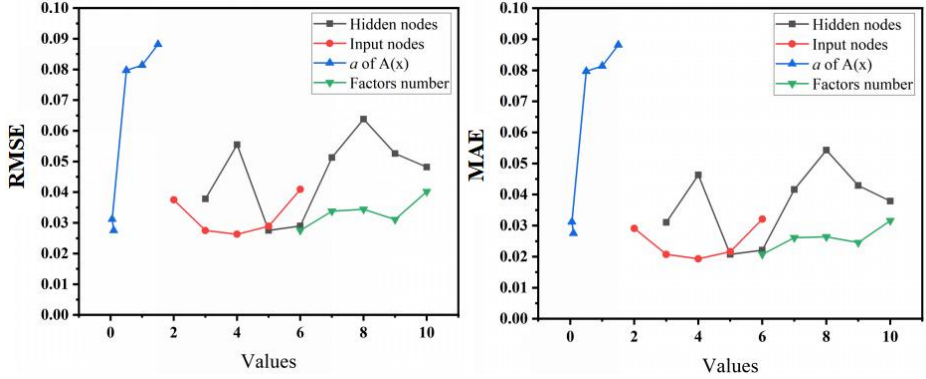

Figure 3: Results of structural optimization for BPNN.

As can be seen from the Figure 3, the optimum hyper-parameters setting is 5 hidden nodes, 4 input nodes, *a* of 0.1 and 6 screened factors.

**3.2 MGHMM for stock timing**

MGHMM is a generative process of an unobservable states sequence by a hidden Markov chain trained and an multivariate observation sequence by the unobservable states sequence. Assume $Q$ is a set containing all the possible states of stock market as $Q = \{q_1, q_2, \cdots, q_N\}$, $I$ is a states sequence as $I = (i_1, i_2, \cdots, i_T)$ and $O$ is the corresponding observation sequence as $O = (o_1, o_2, \cdots, o_T)$, then the initial states distribution is $\vec{\Pi} = [\pi_i]_{N \times 1}$, where $\pi_i = P(i_1 = q_i)$, $i = 1, 2, \cdots, N$; the state transition probability distribution is $A = [a_{ij}]_{N \times N}$, where $a_{ij} = P(i_{t+1} = q_j | i_t = q_i)$, $i = 1, 2, \cdots, N$; $j = 1, 2, \cdots, N$ and the observation probability distribution is $\vec{B} = [b_j(\vec{o})]_{N \times 1}$, in which $b_j(\vec{o})$ obey multivariate Gaussian distribution delineated as equation (7),

$$b_j(\vec{o}) = P(o_t = \vec{o} \mid i_t = q_j)$$
$$\sim N(\vec{o} \mid \vec{\mu}_j, \Sigma_j) = \dfrac{1}{(2\pi)^{n/2}} \cdot \dfrac{1}{|\Sigma_j|^{1/2}} exp\{-\dfrac{1}{2}(\vec{o} - \vec{\mu}_j)^T (\Sigma_j)^{-1}(\vec{o} - \vec{\mu}_j)\} \quad (7)$$

$j = 1, 2, \cdots, N$; $\vec{o} = [v_1, v_2, \cdots, v_n]^T \in \mathbf{R^n}$, wherein $v_i$, $i = 1, 2, \cdots, n$, are the values of *n* observable variables; $\vec{\mu}_j$ and $\Sigma_j$ are the *n*-dimensional mean vector and the *n×n* diagonal covariance matrix corresponding to the *j-th* hidden state severally. The graphical model of MGHMM is portrayed in Figure 4.

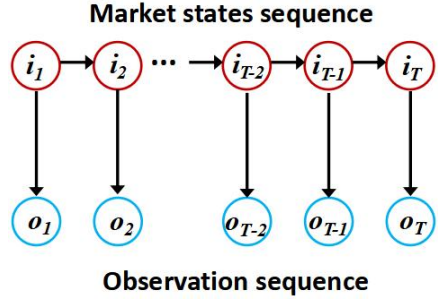

Figure 4: Graphical model of MGHMM.

Moreover, the MGHMM used in this paper is one-order HMM, that is, it obeys homogeneous Markov hypothesis and observation independence hypothesis as usual. Over here, we will monitor the state of CSI 300 index to carry out the stock timing and the number of possible market states is set to be 5. The logarithmic growth rate of financing security balance (LGRFSB), trading volume (TV), daily logarithmic spread of high and low price (DLSHLP), daily logarithmic return (DLR) and five-day logarithmic return (FDLR) of CSI 300 index are chosen to be the observable variables, and the data sequence composed of which require to undergo Box-Cox transformation aiming at the normalization to multivariate Gaussian distribution. Raw data are sourced from Uqer database as well.

Above all, there are two tasks for MGHMM to do given the observation sequence data, one is to solve the model parameter $\lambda = (\overrightarrow{\prod}, A, \vec{\mu}_j, \Sigma_j), j = 1, 2, \cdots, N$ actualizing $\lambda^* = \underset{\lambda}{argmax}\, P(O|\lambda)$ by Baum-Welch algorithm; two is to output the market states sequence $I = (i_1, i_2, \cdots, i_T)$ by Viterbi algorithm, where $i_t = \underset{1 \leq j \leq N}{argmax}\, [P(i_t = q_j|O, \lambda^*)]$, $t = 1, 2, \cdots, T$. During the training process, we will output the corresponding states sequence and total the next trade day's return of every market state inside the states sequence, then 5 market states will be arranged in rank order according to their total.

### 3.3 Fusion strategy of stock timing and picking

#### 3.3.1 Overview of fusion strategy

The quantitative fusion strategy integrates stock picking and timing by PSO-BPNN and MGHMM. To start with screening out six effective factors from 52 stock factors data of CSI 300 index constituent stocks by IC analysis, followed by applying PCA to the data of six factors and leading it to be four dimensions as the input of PSO-BPNN which after training will pick out the candidate stocks for subsequent trading based on the further prediction. Afterwards, MGHMM after training with the CSI 300 index data of its LGRFSB, TV, DLSHLP, DLR and FDLR transformed by Box-Cox approach would output market states sequence, and current market state of which will decide to buy or sell at the next trade day's opening price. Each stock will be bought if the state on the previous trade day ranks in the top two or sold otherwise. The technological procedure diagram is shown in Figure 5.

#### 3.3.2 Trading parameters setting

The effect of a quantitative strategy is not only related to the quality of the practical model, but also closely to the trading operation. The operational details is available in Table 2.

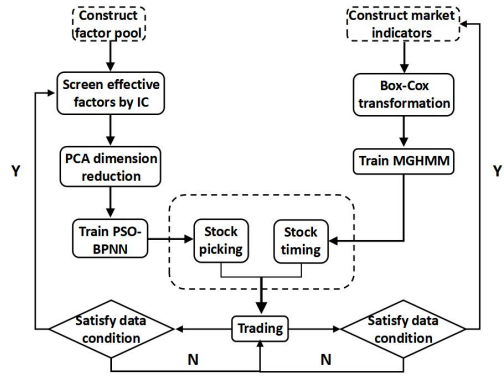

Figure 5: Technological procedure diagram of fusion strategy.

Table 2: The trading operation parameters

| Parameters | Value |
| --- | --- |
| *Backtesting period* | 2020-1-8~2023-9-28 yearly |
| *Initial capital* | 10000000 |
| *Operational cycle* | 1 day |
| *Buy cost* | 0.0003 |
| *Sell cost* | 0.0013 |
| *Slippage* | 0.02 |
| *PSO-BPNN training period* | Past 7 days |
| *MGHMM training period* | Past 40 months |
| *Training cycle* | 1 month |
| *Quantity of candidate stocks* | 3 |
| *Picking cycle* | 7 days |
| *Timing cycle* | 1 day |
| *Buy expense / residual capital* | 30% |
| *Sell proportion* | 100% |

### 3.3.3 Experiment Results

The proposed fusion strategy is compared with Alpha multifactor strategy and Bollinger Bands strategy in Chinese A-Share market. The validation results are provided in Figure 6 and Table 3. It can be seen from Figure 6 that our fusion strategy feckly performed best in terms of return on investment all the past four years except for the latter half of 2020. In the bargain, as demonstrated in Table3 whose upper left, upper right, lower left and lower right refer to 2020, 2021, 2022 and 2023 within the same strategy and criterion, our fusion strategy outperforms Alpha multifactor strategy and Bollinger Bands strategy significantly in criteria of Alpha, Beta, Sharpe Ratio, Information Ratio and Maximum Drawdown, indicating its superb predictive ability as well as towering practicability. It is evident from the Table 3 that our fusion strategy has competence to catch more excessive return under a lower risk as for the investment in stock market.

## 4 CONCLUSION

In this paper we put forward a novel quantitative fusion strategy combining stock picking and timing with PSO-BPNN and MGHMM for the prediction and trading in stock market even the investigation in other financial or eco-nomic domain. The final experiment proved that our fusion strategy is preferable to Alpha multifactor strategy and Bollinger Bands strategy far for forecasting and trading as an investment tool in Chinese A-Share market. Furthermore, the trading operation, the factor pool, PSO parameters, MGHMM parameters and observable variab-les of which, the number of candidate stocks, the frequency of training, picking and timing, etc., all can be adjusted to be suitable for manifold instances and catch financial behavior more accurately as well as more profits.

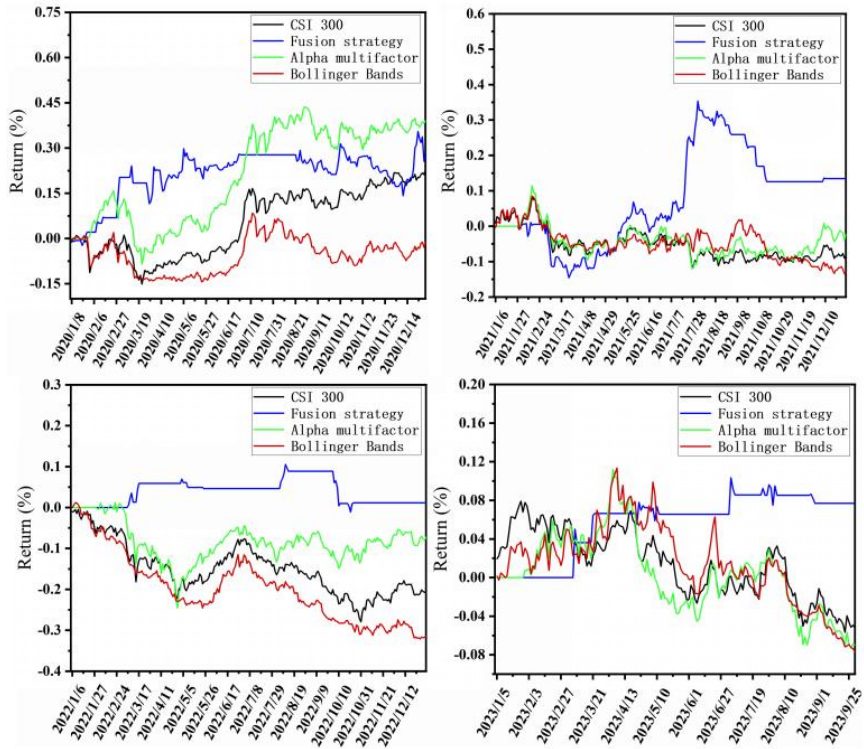

Figure 6: Annual return circumstance over the past four years of different strategies.

Table 3: Annual trading circumstance over the past four years of different strategies

| Strategy | Annualized return (%) | | Alpha (%) | | Beta | | Sharpe Ratio | | Volatility (%) | | Information Ratio | | Maximum Drawdown (%) | | Annualized turnover rate (%) | |
|---|---|---|---|---|---|---|---|---|---|---|---|---|---|---|---|---|
| *Fusion strategy* | 28.8 | 14.1 | 22.4 | 15.8 | 0.14 | 0.43 | 1.07 | 0.49 | 23.7 | 21.7 | 0.12 | 1.00 | 13.2 | 16.8 | 4948.32 | 4700.09 |
| | 1.2 | 10.9 | -0.2 | 7.9 | 0.08 | 0.05 | -0.27 | 0.75 | 8.4 | 9.8 | 1.14 | 1.06 | 10.5 | 2.6 | 1684.27 | 2493.34 |
| *Alpha multifactor* | 41.5 | -2.4 | 20.7 | 4.8 | 0.83 | 0.88 | 1.57 | -0.3 | 24.2 | 19.9 | 0.87 | 0.6 | 21.0 | 21.1 | 768.12 | 793.76 |
| | -7.5 | -9.6 | 10.0 | -4.4 | 0.85 | 0.81 | -0.51 | -0.91 | 21.3 | 14.4 | 1.27 | -0.27 | 25.4 | 16.6 | 773.75 | 639.64 |
| *Bollinger Bands* | -1.3 | 13.2 | -22.2 | -7.6 | 0.84 | 0.75 | -0.21 | -0.85 | 23.1 | 19.7 | -1.72 | -0.33 | 16.1 | 20.3 | 2689.35 | 3097.1 |
| | -32.5 | -10.5 | -22.5 | -5.3 | 0.55 | 0.81 | -2.26 | -0.86 | 16.0 | 16.2 | -1.11 | -0.27 | 32.9 | 17.1 | 2979.48 | 2919.32 |

**APPENDIX**

Table 4: The candidate factors[a]

| | |
|---|---|
| RVI (Relative Volatility Index) | OBV (On Balance Volume) |
| Hurst (Hurst exponent) | ARBR |
| CCI20 (Commodity Channel Index) | CCI5 |
| DEGM (Growth rate of gross income ratio) | BIAS 20 |
| BIAS 5 | REVS 10 |
| ROA (Return on assets) | ROE (Return on equity) |
| RSI (Relative Strength Index) | TotalAssetGrowRate |
| TotalProfitCostRate | VOL120 (Turnover Rate) |
| VOL15 | NetProfitGrowRate |
| NPToTOR (Net profit to total operating revenues) | OperatingProfitGrowRate |
| PB (Price-to-book ratio) | PCF (Price-to-cash-flow ratio) |
| PE (Price-earnings ratio) | PS (Price-to-sales ratio) |
| PSY (Psychological line index) | QuickRatio |
| HBETA (Historical daily beta) | HSIGMA (Historical daily sigma) |
| LCAP (Natural logarithm of total market values) | LFLO (Natural logarithm of float market values) |
| MA5 (Moving average) | MA20 |
| EMA5 (Exponential moving average) | EMA20 |
| MLEV (Market leverage) | NetAssetGrowRate |
| EPS (Earnings per share) | EquityToAsset |
| ETOP (Earnings to price) | FinancialExpenseRate |
| GrossIncomeRatio | CTOP (Cash flow to price) |
| CurrentAssetsRatio | CurrentRatio |
| DAVOL5 | DebtsAssetRatio |
| DilutedEPS (Diluted earnings per share) | AccountsPayablesTRate |
| ARTRate (Accounts receivable turnover rate) | BLEV (Book leverage) |
| BondsPayableToAsset | CashToCurrentLiability |

[a] Sourced from Uqer.